\documentstyle[aps,preprint]{revtex}
\tightenlines
\begin{document}
\title{Thermodynamical properties of metric fluctuations during inflation}
\author{Mauricio Bellini\footnote{E-mail address: mbellini@mdp.edu.ar}}
\address{Instituto de F\'{\i}sica y Matem\'aticas, \\
Universidad Michoacana de San Nicol\'as de Hidalgo,\\
AP: 2-82, (58041) Morelia, Michoac\'an, M\'exico}
\maketitle
\begin{abstract}
I study a thermodynamical approach to scalar metric perturbations
during the inflationary stage.
In the power-law expanding universe here studied, I find
a negative heat capacity as a manifestation of
superexponential growing for the number of states
in super Hubble scales.
The power spectrum depends on the
Gibbons-Hawking and Hagedorn temperatures.
\end{abstract}
\vskip 2cm
During inflation vacuum fluctuations on scales smaller than the Hubble
radius are magnified into classical perturbations of the scalar
field on super Hubble scales. 
The primordial perturbations arise solely from
the zero-point fluctuations of the quantized fields. Although
the region which ultimately expanded to become the observed
universe may have contained excitations above the vacuum, these
excitations would not have any significant effect on the
present state of the universe because a sufficiently large
amount of the inflation would have
redshifted these excitations to inmeasurably long wavelengths\cite{Guth}.
Hence, the density perturbations should be responsible for the
large scale structure formation in the universe\cite{librolinde}.

The scalar perturbations of the metric are related to density
perturbations. These are the spin-zero projections of the graviton,
which only exist in nonvacuum cosmologies. The issue of gauge
invariance becomes critical when attempt to analyze how
the scalar metric perturbations produced in the very early universe
influence a globally flat isotropic and homogeneous universe on super
Hubble scales. This allows us to formulate the problem of the evolution
for the amplitude
of scalar metric perturbations around the Friedmann-
Robertson-Walker (FRW) universe in a coordinate-independent manner at every 
moment in time\cite{Gon}. Since the results do not depend on the gauge, the
perturbed globally flat FRW metric is well described by
\begin{equation}\label{1}
ds^2 = \left(1+2\psi \right) dt^2 - a^2(t) \left(1-2\Phi\right)dx^2,
\end{equation}
where $a$ is the scale factor of the universe and $\psi$, $\Phi$
are the scalar perturbations of the metric. I consider
the particular case where the tensor $T_{ij}$ is diagonal, which
implies that $\Phi=\psi$\cite{Bran}. 
On the other hand I will consider a semiclassical
expansion for the scalar field $\varphi(\vec x,t) = \phi_c(t) +
\phi(\vec x,t)$, with expectation values $\left<0|\varphi |0\right>
=\phi_c(t)$ and $\left<0|\phi|0\right>=0$. Here, $\left.|0\right>$ denotes
the vacuum state\cite{PRD96}. Since $\left<0|\Phi|0\right>=0$, the
expectation value of the metric (\ref{1}) gives the background
metric that describes a flat FRW spacetime.

Linearizing the Einstein equations 
one obtains the system of differential equations for $\phi$ and
$\Phi$\cite{librolinde}
\begin{eqnarray}
&&\ddot\Phi + \left(\frac{\dot a}{a} - \frac{2\ddot\phi_c}{\dot\phi_c}
\right) \dot\Phi - \frac{1}{a^2} \nabla^2\Phi + 2 \left[
\frac{\ddot a}{a} - \left(\frac{\dot a}{a}\right)^2
- \frac{\dot a}{a} \frac{\ddot\phi_c}{\dot\phi_c}\right]\Phi = 0, \label{2}\\
&& \frac{1}{a} \frac{d}{dt} \left( a \Phi\right)_{,\beta} =
\frac{4\pi}{M^2_p} \left(\dot\phi_c \phi\right)_{,\beta}, \label{3}\\
&& \ddot\phi + 3 \frac{\dot a}{a} \dot\phi- \frac{1}{a^2} \nabla^2\phi +
V''(\phi_c) \phi +2 V'(\phi_c) \Phi -4\dot\phi_c\dot\Phi =0. \label{4}
\end{eqnarray}
The dynamics for the background field is given by the equations
\begin{eqnarray}
&& \ddot\phi_c + 3 \frac{\dot a}{a} \dot\phi_c + V'(\phi_c) =0, \label{5}\\
&& \dot\phi_c = - \frac{M^2_p}{4\pi} H'_c(\phi_c),\label{6}
\end{eqnarray}
where $H_c(\phi_c) \equiv \dot a/a$ and $M_p$ is the Planckian mass.
Furthermore, the overdot denotes the time derivative and
the prime represents the derivative with respect to $\varphi$,
evaluated in $\phi_c$ [i.e., $V'(\phi_c) \equiv \left.{d V(\varphi) \over
d\varphi}\right|_{\varphi = \phi_c}$ and $V''(\phi_c) \equiv
\left.{d^2 V(\varphi) \over d\varphi^2}\right|_{\varphi = \phi_c}$ ].

The eq. (\ref{2}) can be simplified by means of the map $h = e^{1/2 \int
\left({\dot a\over a} - {2 \ddot\phi_c \over \dot\phi_c} \right) dt}
\Phi$, and one obtains a Klein-Gordon equation for the redefined scalar
metric fluctuations $h$\cite{Bellini}

\begin{eqnarray}
\ddot h & - & \frac{1}{a^2} \nabla^2 h - \left\{ \frac{1}{4}
\left(\frac{ \dot a}{a} - \frac{2 \ddot\phi_c}{\dot\phi_c}\right)^2+
\frac{1}{2} \left(\frac{\ddot a - \dot a^2}{a^2}-
\frac{2\frac{d}{dt}\left(\ddot\phi_c\dot\phi_c\right) - 4 \ddot\phi^2_c}{
\dot\phi^2_c} \right) \right. \nonumber \\
& - & \left.2\left[ \frac{\ddot a}{a} - \left(\frac{\dot a}{a}\right)^2 -
\frac{\dot a}{a} \frac{\ddot\phi_c}{\dot\phi_c} \right]\right\} h =0.
\label{7}
\end{eqnarray}
This field can be written as a Fourier expansion in terms of the modes
$h_k = e^{i\vec k. \vec x} \xi_k(t)$
\begin{equation}\label{8}
h(\vec x,t) = \frac{1}{(2\pi)^{3/2}} {\Large \int} d^3k \left[
a_k h_k + a^{\dagger}_k h^*_k \right],
\end{equation}
where $\xi_k(t)$ are the time dependent modes and $(a_k,a^{\dagger}_k)$
are the annihilation and creation operators which complies with
the commutation relations $\left[a_{\vec k}, a^{\dagger}_{\vec k'}\right]=
\delta^{(3)}(\vec k - \vec k')$ and $\left[ a_{\vec k},a_{\vec k'}\right]
= \left[a^{\dagger}_{\vec k} , a^{\dagger}_{\vec k'} \right] =0$.
The commutation relation for $h$ and $\dot h$ is
$[h,\dot h]=i \delta^{(3)}(\vec x-\vec x')$ which
implies that $\xi_k \dot\xi^*_k - \dot\xi_k \xi^*_k =i$, for any time.
Replacing eq. (\ref{8}) in (\ref{7}) one obtains the equation of
motion for $\xi_k(t)$
\begin{eqnarray}
\ddot\xi_k &+& \left\{ \frac{k^2}{a^2} +2 \left[ \frac{\ddot a}{a}
-\left(\frac{\dot a}{a}\right)^2 - \frac{\dot a}{a}
\frac{\ddot\phi_c}{\dot\phi_c}
\right] - \frac{1}{4}\left(\frac{\dot a}{a}
- 2 \frac{\ddot\phi_c}{\dot\phi_c}
\right)^2 \right. \nonumber \\
&-& \left.\frac{1}{2} \left( \frac{\ddot a a - \dot a^2}{a^2} - \frac{
2\frac{d}{dt}\left(\ddot\phi_c \dot\phi_c\right) - 4 
\ddot\phi^2_c}{\dot\phi^2_c}\right)\right\}\xi_k =0,\label{9}
\end{eqnarray}
which can be written in a simplified manner as 
$\ddot\xi_k +\omega^2_k(t) \xi_k=0$, where
\begin{equation}\label{10}
\omega^2_k = a^{-2} [k^2 - k^2_0(t)],
\end{equation} 
is the squared frequency with wavenumber $k$ 
and the effective parameter of mass is
\begin{equation}\label{11}
\mu^2(t) = \frac{k^2_0(t)}{a^2} = 
\frac{1}{4} \left(\frac{\dot a}{a} -\frac{2\ddot\phi_c}{\dot\phi_c}\right)^2
+\frac{1}{2} \left( \frac{\ddot a a - \dot a^2}{a^2} - \frac{
2\frac{d}{dt}\left(\ddot\phi_c \dot\phi_c\right) - 4 \ddot\phi^2_c}{
\dot\phi^2_c} \right)-
2 \left[ \frac{\ddot a}{a} - \left(\frac{\dot a}{a}\right)^2
-\frac{\dot a}{a} \frac{\ddot\phi_c}{\dot\phi_c}\right].
\end{equation}
Here, $k_0(t)$ is the time dependent wavenumber which separates
the infrared (IR) (with $k \ll k_0$) and ultraviolet (UV) (with
$k \gg k_0$) sectors.
The IR sector describes the super Hubble dynamics for the metric
fluctuations which are responsible for gravitational effects
on cosmological scales.

The issue of the increasing number of degrees of freedom for matter field
fluctuations in the IR sector in the framework of thermodynamics
was studied in another work\cite{pr2000}. 
The main aim of this work is the study of thermodynamical properties
for scalar metric fluctuations on super Hubble scales. 
To make a thermodynamical description for these fluctuations we can write
the partition function $Z(\beta)$ on super Hubble scales
\begin{equation}\label{12}
Z(\beta) \simeq {\Large\int}^{\epsilon k_0}_{k=0}
\frac{d^3k}{(2\pi)^{3/2}} \  e^{-\beta\omega_k(t)}
={\Large\int}^{\epsilon k_0}_{k=0} d\omega_k \  \rho(\omega_k)
e^{-\beta \omega_k},
\end{equation}
where $\beta^{-1}$ is related to the background ``temperature''. 
The squared frequency with cut-off wavenumber $\epsilon k_0$ is given
by $\omega^2_{\epsilon k_0} = - a^{-2} \left[ k^2_0 (1-\epsilon^2)\right]$,
where $\epsilon$ is a dimensionless parameter given by $k/k_0 \ll 1$. In
the semiclassical limit the frequency $\omega_k$ plays the role
of the energy for each mode with wavenumber $k$. 
In the IR sector the wavenumbers are very
small with respect to $k_0$ ($k \ll k_0$), and the frequency
$\omega_k$ is imaginary pure ($\omega_k = \pm i |\omega_k|$).
The function $\rho(\omega_k)$ gives the density of states with frequency
$\omega_k$ 
\begin{equation}\label{13}
\rho(\omega_k) = \frac{1}{(2\pi)^3} \left|\frac{d^3k}{d\omega_k}\right|=
\frac{k^2}{2\pi^2}\left|\frac{dk}{d\omega_k}\right|,
\end{equation}
where $\left|{d^3k \over d\omega_k}\right|$ is the Jacobian of the
transformation such that
$\left|{d\omega_k\over dk}\right| 
= {\left[k^2_0 + a^2 \omega^2_k\right]^{1/2}\over a^2 |\omega_k|}$,
where $|\omega_k| = [\omega_k \omega^*_k]^{1/2}$ and the asterisk
denotes the complex conjugate. Hence, the density of states with
frequency $\omega_k$ can be written as
\begin{equation}\label{15}
\rho(\omega_k) = \frac{1}{2\pi^2} 
\left[k^2_0 + a^2(t) \omega^2_k\right]^{1/2}
|\omega_k| a^2(t).
\end{equation}
Energy added to a system can go either into increasing
the energy of existing states or into creating new
states. In the case of the IR gauge-invariant metric fluctuations
new and new states are created from the UV sector
during inflation. These fluctuations can be written as a Fourier expansion
taking into account only the modes with $k \ll k_0$
\begin{equation}
h_{cg} = \frac{1}{(2\pi)^{3/2}} {\Large\int} d^3k \  \theta(\epsilon k_0 - k)
\left[ a_k h_k + a^{\dagger}_k h^*_k\right],
\end{equation}
where $\epsilon \ll 1$ is a dimensionless parameter and
$k_0$ is the time dependent wavenumber that separates the IR and UV
sectors.
The inverse of ``temperature'' and the heat capacity are given by
\begin{eqnarray}
\beta &=& \left. \frac{\partial {\rm ln}[\rho]}{\partial \omega_k}\right|_{
k=\epsilon k_0}, \label{16}\\
C_V &=& -(\beta \beta^*) \left.\left[\frac{\partial^2 {\rm ln}[\rho]}{\partial\omega^2_k}
\right]\right|_{k=\epsilon k_0}.\label{17}
\end{eqnarray}

The quantum nature of the metric fluctuations
(in the UV sector) is the motivation for which $\beta$
is imaginary pure. 
If the second derivative in eq. (\ref{17}) is positive, hence
$C_V$ becomes negative and the density of states rises superexponentially.
Systems with negative heat capacity are thermodynamically unstable. They
are placed in contact with a heat bath and will experience runway heating
or cooling. If the density of states grows exponentially, an inflow
of energy at the Hagedorn temperature\cite{Hagedorn} 
goes entirely into producing new
states, leaving the temperature constant. If the density of states
grows superexponentially, the process is similar, but the production
of new states is so copious that the inflow of energy actually
drives the temperature down. In our case the Hagedorn temperature
is not a true temperature. It takes into account the energy
of each mode $h_k$.

To simplify de notation, I will denote $\omega_{\epsilon k_0}$ as $\omega$.
During inflation the heat capacity for IR gauge-invariant scalar metric
perturbations is given by
\begin{equation}\label{18}
C_V = \frac{- \mu^4 \left(\omega^2 \mu^2 +\mu^4 + 2\omega^4\right)}{
\omega^4 \left(\mu^2+\omega^2\right)^4},
\end{equation}
where $\mu^2=k^2_0/a^2 > 0$. The inverse of the effective ``temperature'' is
\begin{equation}\label{19}
\beta \simeq \mp i \  \frac{\mu^2}{|\omega| \left(\mu^2+\omega^2\right)}.
\end{equation}
Note that $\beta$ describes the environment of the IR sector.
If $C_V >0$, the system distributes its
energy in the existent states. The situation $C_V <0$
describes a system which increments very rapidly the number of states.

Now I consider the case of a power-law expanding universe for
which the scale factor evolves as $a(t) \propto \left(t/t_0\right)^p$
and the Hubble parameter is $H_c[\phi_c(t)]=p/t$. The background
field is given by $\phi_c(t) = \phi^{(0)}_c-m \  {\rm ln}[(t/t_0)p]$,
where $\phi^{(0)}_c = \phi_c(t=t_0)$ and $t\ge t_0$. Furthermore,
$m \simeq (10^{-4} - 10^{-6}) \  M_p$ is the mass of the inflaton
field.
In this particular case $h(\vec x,t) = t^{({p\over 2} +1)} \  \Phi(\vec x,t)$
and $\mu^2 = {k^2_0 \over a^2}$ being given by
\begin{equation}
\mu^2(t) =  M^2 \  t^{-2},
\end{equation}
where $M^2=(p^2+4)/4$. The density of states and $\beta$ are
\begin{eqnarray}
\rho(\omega_k) & \simeq & \frac{|\omega_k|}{2\pi^2} 
t^{3p}
\left[ M^2 \  t^{-2} + \omega^2_k\right]^{1/2}, \\
\beta(t) & \simeq & \mp {\rm i} \frac{
t}{\epsilon^2 M}.
\end{eqnarray}
Finally, the heat capacity that results is
\begin{equation}
C_V \simeq - 2 \left(\beta\beta^*\right)^2 \simeq
-2  \frac{t^4}{\epsilon^8 M^4},
\end{equation}
which is negative but its absolute value increases as
$t^{4}$. This is a manifestation of the superexponential growth
of the number of degrees of freedom for the IR scalar metric fluctuations
in a power-law expanding inflationary universe.
The Hagedorn temperature $(\beta\beta^*)^{-1/2}$ go asymptotically
to zero as $t^{-1}$.
Both facts, $\dot C_V <0$ and 
$\left.[\beta\beta^*]^{-1/2}\right|_{t\rightarrow
\infty} \rightarrow 0$, are consequence of the instability of the
IR sector during inflation. Such an instability is due to the interaction
of the inflaton field, which manifests itself in the temporal dependence of 
the effective parameter of mass $\mu(t)\sim t^{-1}$.

If $(\beta\beta^*)^{-1/2}$ is the zero mode ``temperature'' (or
background temperature), the squared metric perturbations when the horizon
entry will be given by 
\begin{equation}
\left<\Phi^2_{cg}\right>_{IR} = {\Large\int}^{\epsilon k_0}_{k=0}
\frac{dk}{k} {\cal P}_{\Phi_{cg}}(t) \simeq \left. t^{-(p+2)}
{\Large\int}_0^{\epsilon k_0} \frac{d^3k}{(2\pi)^3} \  \left[\xi_{k=0}(t)
\right]^2\right|_{t_*},
\end{equation}
where $t_*$ is the time when the horizon entry and $\xi_{k=0}$ is the
solution of $\ddot\xi_0 - \left(\frac{\epsilon^2}{\beta\beta^*}\right)\xi_0=0$.
The asymptotic solution for this equation is 
\begin{equation}
\xi_0(t) \simeq c_1 \  t^{\frac{1}{2}\left[1+ \sqrt{1+4M^2}\right]},
\end{equation}
In other words
(for $p > 3.04$ needed to power-law inflation takes place), 
the power spectrum ${\cal P}_{\Phi_{cg}}(t_*) \simeq
\left.2 \epsilon^2 \left(t^{3p + \sqrt{5+p^2}-1}\right)_{t_*}
{T^2_H \over T^2_{GH}(t)}\right|_{t_*} \  k^3_* $ 
is a function of the Gibbons-Hawking and 
Hagedorn temperatures, given respectively by $T_{GH}=H_0/(2\pi)$
and $T_H=(\beta\beta^*)^{-1/2}$.
Here, $T_{GH}$ is the temperature of the primordial horizon
(with size $H^{-1}_0$) and $T_H(t_*)$ is the temperature
related to the cosmic horizon at the moment
of horizon-crossing [which has a size $H^{-1}(t_*)$].

To summarize, super Hubble metric fluctuations with negative heat capacity during
power-law inflation describes superexponential growth of the number of
states, which is a characteristic of nonequilibrium thermodynamical systems.
Rather an increasing energy density, an increasing of $|C_V|$ (for 
$C_V <0$), gives a superproduction of the number of degrees of freedom
in the infrared sector. Notice that the heat capacity in the
model here studied decreases very rapidly. 
The imaginary nature of $\beta$ must be understood as a consequence
of the quantum nature of the metric fluctuations in the ultraviolet
sector, which plays the role of infrared's environment in the
coarse-grained field $h_{cg}$-representation here developed. \\
\vskip .2cm
\centerline{\bf ACKNOWLEDGMENTS}
\vskip .2cm
\noindent
I would like to acknowledge CONACYT (M\'exico) and CIC of Universidad
Michoacana for financial support in the form of a research grant.\\

\end{document}